\documentclass[prl,aps,amssymb,nofootinbib]{revtex4}
\usepackage{graphicx}
\usepackage{epsf}
\begin{document}
\title{Effective field theory description of the higher dimensional quantum Hall liquid}
\author{B. Andrei Bernevig$^{2}$, Chyh-Hong Chern$^{1}$, Jiang-Ping Hu$^{1}$, Nicolaos Toumbas$^{3}$ and Shou-Cheng Zhang$^{1}$}

\affiliation{$^{1}$Department of Physics, McCullough Building,
Stanford University, Stanford  CA 94305-4045 \\ $^{2}$ Department of Physics, Massachusetts Institute of Technology, Cambridge, MA 02139 \\
$^{3}$ Jefferson Physical Laboratory, Harvard University,
Cambridge, MA 02138}

\begin{abstract}
\par~\par
We derive an effective topological field theory model of the
four dimensional quantum Hall liquid state recently constructed
by Zhang and Hu. Using a generalization of the flux attachment
transformation, the effective field theory can be formulated
as a $U(1)$ Chern-Simons theory over the total configuration
space $CP_3$, or as a $SU(2)$ Chern-Simons theory over $S^4$.
The new quantum Hall liquid supports various types of topological
excitations, including the 0-brane (particles), the 2-brane (membranes)
and the 4-brane. There is a topological phase interaction among
the membranes which generalizes the concept of fractional
statistics.

\end{abstract}
%\pacs{74.20-z, 11.30.Ly, 74.25.Ha, 74.25.Jb}
\maketitle \pagebreak

\section{Introduction}
Recently, a higher dimensional generalization of the quantum Hall
effect was constructed by two of us (ZH)\cite{zhang2001}. This
incompressible liquid was defined on a 4 dimensional spherical
surface with a $SU(2)$ monopole at its
center\cite{belavin1975,jackiw1976,Yang}. The fermionic particles
carry a $SU(2)$ quantum number $I$, which scales with the radius
$R$ as $I\sim R^2$. This quantum liquid state shares many
properties with its 2 dimensional counterpart\cite{laughlin1983}.
The ground state is separated from all excited states by a finite
energy gap, and the density correlation functions decay
gaussianly. Fractional quantum Hall states can also be
constructed, and they support fractionally charged quasi-particle
excitations. The full spectrum of the boundary excitations of this
quantum liquid is still not fully understood, but a partial
analysis reveals collective excitations with all relativistic
helicities\cite{zhang2001,hu2001}. More recently,
Fabinger\cite{fabinger} found a realization of this system within
string theory. Sparling\cite{sparling} found a deep connection
between this system and the twistor theory. Karabali and
Nair\cite{nair} investigated the generalization of the quantum
Hall effect on the $CP_n$ manifolds. Ho and Ramgoolam
\cite{ramgoolam} obtained the matrix descriptions of even
dimensional fuzzy spherical branes $S^{2k}$ in Matrix Theory.
Chen, Hou and Hou\cite{hou} discussed the relationship between the
4 dimensional quantum Hall liquid and the non-commutative geometry
on $S^4$.

Emergence of the Chern-Simons (CS) gauge structure in the two
dimensional fractional quantum Hall effect (FQHE) was an exciting
development in condensed matter physics in recent
years\cite{zhang1989,zhang1992,susskind}. The
Chern-Simons-Landau-Ginzburg (CSLG) theory of the FQHE describes
the long wave length physics of the incompressible quantum Hall
liquid in terms of a topological field theory, which to the
leading order is independent of the space-time metric. $1+1$
dimensional chiral relativistic dynamics emerges when the CSLG
theory is restricted to the edge of the quantum Hall
liquid\cite{wen1990,stone1990}. It would be highly desirable to
see how this elegant connection between the microscopic wave
function and the topological field theory can also be generalized
to higher dimensions.

In this work we focus on the effective topological field theory of
the quantum liquid state constructed by ZH. The key step is to
generalize the concept of flux attachment
transformation\cite{zhang1989,zhang1992} to higher dimensions, and
to use the concept of non-commutative geometry\cite{douglas2001}
to relate theories defined in different dimensions. We find two
equivalent CS theories, an abelian CS theory in 6+1 dimensions and
a $SU(2)$ non-abelian CS theory in 4+1 dimensions. The quantum
liquid constructed by ZH has orbital degrees of freedom scaling
with $R^4$ and internal isospin degree of freedom scaling with
$R^2$. Therefore, the total configuration space is 6
dimensional\cite{zhang2001}. This can also be seen from the second
Hopf map $S^7\rightarrow S^4$, which was the central mathematical
construct used by ZH. $S^7$ describes the combined orbital and
isospin degrees of freedom. However, since quantum mechanics is
based on $U(1)$ projective representations, the actual
configuration space is $S^7/U(1)=CP_3$, or the 3 complex
dimensional (6 real dimensional) projective space. Our 6+1
dimensional CS theory is naturally defined over the $CP_3\times
{\cal R}$ manifold, where the 6 volume form on $CP_3$ plays the
role of the generalized background flux. Since the $CP_3$ manifold
locally decomposes as $S^4\times S^2$, we can also arrive at a
$4+1$ dimensional continuum field theory by treating $S^2$ as a
"fuzzy sphere", with discrete matrix model degrees of freedom. By
this procedure, we arrive at the equivalent $SU(2)$ non-abelian CS
gauge field theory in $4+1$ dimensions, where particles with
$SU(2)$ internal isospin degrees of freedom are attached to the
$SU(2)$ instanton density over $S^4$.

Besides the quasi-particle like elementary excitations, we show here that
the quantum liquid
constructed by ZH also supports topologically stable extended objects,
namely the membrane (2-brane) and the 4-brane. The topological action
for these extended objects can be derived exactly. A particularly interesting
aspect of the membranes is that they have non-trivial phase interactions\cite{Wu+Zee,tze}
which is a direct generalization of the fractional statistics carried by the
Laughlin quasi-particles in the 2D quantum Hall liquid.

\section{Single particle Lagrangian}

The 2D quantum Hall effect arises when a two-dimensional electron
gas is subjected to a strong magnetic field perpendicular to its
plane.  It can also arise when electrons move on a two sphere
$S^2$ with a $U(1)$ magnetic monopole at the center
\cite{Haldane}. The two systems can be related to each other by a
conformal transformation, namely, the inverse stereographic
projection from $S^{2}$ to $R^{2}$, with all points at infinity
identified with the south pole. Much of the spherical physics
relies on the topological properties the first Hopf map, $S^{3}\to
S^{2}$. To exhibit the map, we introduce a two component complex
spinor $u$ with $\bar{u}u=1$ and set
\begin{eqnarray}
& & u=\left( \begin{array}{c} u_{1} \\ u_{2} \end{array} \right)=\left( \begin{array}{c} \cos\frac{\theta}{2}e^{i\frac{\phi}{2}+i\frac{\chi}{2}} \\
\sin\frac{\theta}{2}e^{i\frac{\phi}{2}-i\frac{\chi}{2}}\end{array}
\right) \label{z}
\end{eqnarray}
where $\theta$ is in $[0,\pi]$; $\phi$ and $\chi$ are in
$[0,2\pi)$. The $1^{st}$ Hopf map is defined as
\begin{eqnarray}
\frac{n_{i}}{r}=\bar{u}\sigma_{i}u, \ \ \ i=1,2,3 \label{1sthopf}
\end{eqnarray}
where $\sigma_{i}$'s are the three Pauli matrices, and
$n^{2}_{i}=r^{2}$ is manifestly satisfied.  Here $n_i$ is the
coordinate of $S^{2}$ with radius r.  The Hopf fibration defines
the principal U(1) bundle over $S^{2}$. The Dirac quantization
condition on the sphere gives $eg=I$, where $e$ and $g$ are the
electric and magnetic charges respectively. $I$ is quantized to be
an integer or a half-integer. We are interested in the limit of
infinite $I$ and $r$, such that the ratio $I/r^2$ is fixed. The
many-body electronic system can be thought of as an incompressible
liquid on $S^{2}$.

In the strong magnetic field limit, or in the limit of vanishing
kinetic mass $m\rightarrow 0$, the single particle Lagrangian
contains only a first order time derivative. It is in fact nothing
but the Berry's phase of the $u$ spinors. Since $S^3/U(1)=S^2$,
the $U(1)$ Berry's phase over the base space $S^2$ can be simply
expressed in the the following form
\begin{eqnarray}
L=2iI\bar{u}\frac{d}{dt}u=iITr(\sigma_{3}s^{-1}(x)\dot{s}(x))
\label{2dlagrangian}
\end{eqnarray}
Here $I$ is chosen to label the spinor representation of $SU(2)$,
and $s(x)$ is an element of SU(2).  It can be expressed as
\begin{eqnarray}
& & s(x)\!=\!\left(\begin{array}{cc} u_{1} & u_{2}^{*} \\ -u_{2} &
u_{1}^{*}\end{array} \right) \label{su2element}
%\nonumber & & s(x) \!=\! \left(
%\begin{array}{cc}
%cos\frac{\theta}{2}e^{i\frac{\phi}{2}}e^{i\frac{\chi}{2}}   & sin\frac{\theta}{2}e^{-i\frac{\phi}{2}}e^{i\frac{\chi}{2}}  \\
%-sin\frac{\theta}{2}e^{i\frac{\phi}{2}}e^{-i\frac{\chi}{2}} &
%cos\frac{\theta}{2}e^{-i\frac{\phi}{2}}e^{-i\frac{\chi}{2}}
%\end{array} \right).
\end{eqnarray}
The second form in (\ref{2dlagrangian}) is due to
Balachandran\cite{balabook}. In the gauge where
$\dot{\phi}=-\dot{\chi}$, (\ref{2dlagrangian}) becomes
\begin{eqnarray}
L=I\varepsilon_{3ij}\frac{n_{i}\dot{n_{j}}}{r(r+n_{3})}
\end{eqnarray}
This is the Lagrangian of a particle moving on $S^2$, interacting
with a $U(1)$ monopole gauge potential. The Dirac string of this
monopole potential is located at the south pole in this gauge.
This gauge will be used throughout the paper unless otherwise is
stated.

In a higher dimensional generalization, ZH\cite{zhang2001}
considered fermions moving on a 4-sphere interacting with an
$SU(2)$ magnetic monopole at the center
\cite{belavin1975,jackiw1976,Yang}. Yang \cite{Yang} has shown
that the $SU(2)$ monopole with $\pm 1$ topological charge is
$SO(5)$ invariant. The field strength is self-dual for the
topological charge $+1$ and anti-self-dual for the opposite case.
The underlying algebraic structure of this system is the second
Hopf fibration, $S^{7}\to S^{4}$. Let $\Psi$ be a 4-component
complex spinor with $\bar{\Psi}\Psi=1$. The second Hopf map is
given by
\begin{eqnarray}
\frac{X_{a}}{R}=\bar{\Psi}\Gamma_{a}\Psi \label{2ndhopf}
\end{eqnarray}
Here $\Gamma_{a}$'s are the five Dirac Gamma matrices, satisfying
the Clifford algebra $\{\Gamma_{a},\Gamma_{b}\}=2\delta_{ab}$. It
is easy to see that $X_a^2=R^2$ follows from the normalization
condition $\bar{\Psi}\Psi=1$. Since $\Psi$ is a $4$ component
complex spinor, the normalization condition defines a $7$ sphere
$S^7$ embedded in $8$ dimensional Euclidean space. On the other
hand, $X_a^2=R^2$ defines a $4$ sphere $S^4$ with radius $R$
embedded in $5$ dimensional Euclidean space. Therefore, Eq.
(\ref{2ndhopf}) defines a mapping from $S^7$ to $S^4$. $S^7$ can
be viewed as a principal $SU(2)$ bundle over $S^{4}$. It can also
be viewed as a $U(1)$ bundle over $S^7/U(1)=CP_3$, namely the
complex projective space with (complex) dimension three.

To generalize (\ref{2dlagrangian}), we consider the Berry's phase
for the $\Psi$ spinor. The $SU(2)$ non-abelian Berry's phase over
the base space $S^4$ has been computed by Demler and Zhang in ref.
\cite{demler1999}. Here we shall find it useful to compute the
$U(1)$ abelian Berry's phase over $CP_3$. Since $CP_3$ can be
defined as the space of the $\Psi$ spinors up to an overall $U(1)$
phase factor, the $U(1)$ Berry's phase Lagrangian is simply given
by
\begin{eqnarray}
L=2iI\Psi^{\dagger}\frac{d}{dt}\Psi \label{splagrangian}
\end{eqnarray}
We take $\Psi$ to be the solution of the Hopf equation, which is
given by $\frac{X_{a}}{R}\Gamma_{a}\Psi=\Psi$.  Then, $\Psi$ is
solved as
\begin{eqnarray}
\Psi \!=\! \left(
\begin{array}{c}
\Psi_{1} \\
\Psi_{2} \\
\Psi_{3} \\
\Psi_{4}
\end{array} \right)
\!=\! \left(
\begin{array}{c}
\sqrt{\frac{R+X_{5}}{2R}}\left(\begin{array}{l} u_{1} \\ u_{2} \end{array} \right)\\
\frac{1}{\sqrt{2R(R+X_{5})}}(X_{4}-iX_{i}\sigma^{i})\left(\begin{array}{l}
u_{1} \\ u_{2} \end{array} \right) \end{array} \right) \label{Psi}
\end{eqnarray}
where
\begin{eqnarray}
& & u=\left( \begin{array}{c} u_{1} \\ u_{2} \end{array} \right)
=\left( \begin{array}{c} \sqrt{\frac{r+n_3}{2r}} \\
\frac{n_1+in_2}{\sqrt{2r(r+n_3)}}\end{array} \right) \label{u}
\end{eqnarray}
We can write the action more explicitly as
\begin{eqnarray} \nonumber
& &S=2iI\int dt \bar\Psi
\partial_t\Psi=-\int dt
(I\varepsilon_{3ij}\frac{n_{i}\dot{n_{j}}}{r(r+n_{3})}+\frac{I}{R(R+X_{5})}\eta^{i}_{\mu\nu}\frac{n_{i}}{r}X_{\mu}\dot{X_{\nu}})
 \\ \nonumber & &=\int dt (A_a(X,n) \partial_t X_a + A_i(X,n)
\partial_t n_i)
\\ & &\equiv \int dt g^{AB}A_A \partial_t X_B \label{single}
\end{eqnarray}
where
$\eta^{i}_{\mu\nu}=\varepsilon_{i\mu\nu4}+\delta_{i\mu}\delta_{4\nu}-\delta_{i\nu}\delta_{4\mu}$
is the 't Hooft symbol. $X_B$ denotes the local coordinate of the
$CP_3$ manifold. From the parameterization of $\Psi$ and $u$ given
in (\ref{Psi}) and (\ref{u}), we see explicitly that the $CP_3$
manifold locally decomposes as $S^4\times S^2$. Index conventions
and the explicit form of the metric tensor $g_{AB}$ over $CP_3$
are given in the Appendix. The quantization of the Lagrangian will
be determined later. From (\ref{single}), we can read out the
$U(1)$ gauge connection over the $CP_3$ manifold:
\begin{eqnarray}
A_{\mu} &=& \frac{I}{R(R+X_{5})}\eta^{i}_{\mu\nu}\frac{n_{i}}{r}X_{\nu}, \ \ \ A_{5}=0  \\
A_{i} &=& \frac{I}{r(r+n_{3})}\varepsilon_{3ij}n_{j}
\label{connection}
\end{eqnarray}
The gauge potential is defined only patch by patch. Here the gauge
potential is defined over the north pole region $X_5\approx R$ and
$n_3\approx r$. A similar gauge potential can be defined over the
south pole region. In the overlap of patches, the gauge potentials
defined in each patch differ by a $U(1)$ gauge transformation. The
gauge potential also satisfies the following transversality
conditions
\begin{eqnarray}
X_{a}A_{a}=0, \ \ n_{i}A_{i}=0
\end{eqnarray}

\section{Derivation of equations of motion}
The field strength of the $U(1)$ gauge potential, $F_{\lambda\tau}
(\lambda,\tau=\{a,i\})$, is given by
\begin{eqnarray}
F_{\lambda\tau}=\partial_{\lambda}A_{\tau}-\partial_{\tau}A_{\lambda}
\end{eqnarray}
Since we are using a redundant set of coordinates here, we need to
be careful and differentiate $R$ and $r$, using $\partial
R/\partial X_a=\frac{X_{a}}{R}$ and $\partial r/\partial
n_i=\frac{n_{i}}{r}$. The resulting field strength is given by
\begin{eqnarray}
F_{\mu\nu} & = &
\frac{-2I}{R(R+X_{5})}\eta^{i}_{\mu\nu}\frac{n_{i}}{r}+
\frac{2R+X_{5}}{R^{2}(R+X_{5})}(A_{\mu}X_{\nu}-A_{\nu}X_{\mu})
 \nonumber \\
F_{\mu i} & = &
\frac{-I}{rR(R+X_{5})}\eta^{i}_{\mu\nu}X_{\nu}+\frac{I}{rR(R+X_{5})}
\eta^{k}_{\mu\nu}\frac{n_{k}}{r}\frac{n_{i}}{r}X_{\nu}
\nonumber \\
F_{5i} & = & 0, ~~~ F_{\mu5}=
\frac{I}{R^{3}}\eta^{i}_{\mu\tau}\frac{n_{i}}{r}X_{\tau}, ~~~
F_{ij}=\frac{I}{r^{3}}\varepsilon_{ijk}n_{k} \label{F}
\end{eqnarray}
 and the following equations are satisfied
\begin{eqnarray}
X_{a}F_{ab}=0, \ \ n_{i}F_{ij}=0, \ \ F_{ai}n_{i}=0, \ \
F_{ai}X_{a}=0
\end{eqnarray}
The equations of motion can be derived from the Lagrangian
(\ref{single}) together with an external potential $V(n,X)$. Using
Lagrange multipliers for the constraints, $X_a^2=R^2$ and
$n_i^2=r^2$, and taking the variation of $X_{b}$ and $n_k$
respectively, we obtain the following two equations of motion
\footnote{For convenience, we show the results with
$I=\frac{1}{2}$.  The general expressions follow easily.}
\begin{eqnarray} \nonumber
& &F_{ba}\dot{X}_{a}+\frac{X_{a}}{R^{2}}\frac{\partial V}{\partial
X_{a}}X_{b}-\frac{\partial V}{\partial X_{b}}+F_{bi}\dot{n}_{i}=0
\\
&
&F_{ki}\dot{n}_{i}+F_{ka}\dot{X}_{a}+\frac{n_{i}}{r^{2}}\frac{\partial
V}{\partial n_{i}}n_{k}-\frac{\partial V}{\partial n_{k}}=0 \
\end{eqnarray}
The above equations can be simplified to
\begin{eqnarray}
& &
(F_{ba}-4r^2F_{bi}F_{ik}F_{ka})\dot{X}_{a}=-\frac{X_{a}}{R^{2}}\frac{\partial
V}{\partial X_{a}}X_{b}+\frac{\partial V}{\partial X_{b}}-4r^4
F_{bj}F_{jk}\frac{\partial V}{\partial n_{k}}
\nonumber \\
 & &\dot{n}_{j}=4r^{4}F_{jk}\frac{\partial V}{\partial
n_{k}}-4r^{4}F_{jk}F_{k\mu}\dot{X}_{\mu} \label{eomu1}
\end{eqnarray}

For most cases we shall consider, the potential $V(n,X)$ does not
depend on the isospin coordinates, {\it i.e.} $\partial V/\partial
n=0$ on the right hand side of (\ref{eomu1}).  The first term at
the right hand side of the second equation in (\ref{eomu1}) is
zero. Recall that the $SU(2)$ matrix valued gauge potential ${\bf
A}_a$ and field strength ${\bf F}_{ab}=\partial_a {\bf A}_b -
\partial_b {\bf A}_a + [{\bf A}_a, {\bf A}_b]$ are explicitly
given by\cite{zhang2001}
\begin{eqnarray} \nonumber
{\bf A}_{\mu} &=& \frac{-i}{R(R+X_{5})}\eta^{i}_{\mu\nu}{\bf
I}_{i}X_{\nu}, \ \ \ {\bf A}_{5}=0 \\ \nonumber{\bf F}_{\mu\nu}&=&
\frac{1}{R^{2}}(X_{\nu}{\bf
A}_{\mu}-X_{\mu}{\bf A}_{\nu}+i\eta^{i}_{\mu\nu}{\bf I}_{i}) \\
{\bf F}_{5\mu}&=&\partial_{5}{\bf
A}_{\mu}=-\frac{R+X_{5}}{R^{2}}{\bf A}_{\mu}
\label{su2strength}
\end{eqnarray}
The Heisenberg operator equations of motion in the large field
($m\rightarrow 0$) limit are given by
\begin{eqnarray}
\dot{X}_{b}{\bf F}_{ab} & = &i(\frac{\partial V}{\partial
X_{a}}-\frac{X_{a}X_{b}}{R^{2}}\frac{\partial V}{\partial X_{b}})
\nonumber \\
\dot{{\bf I}}_{i}&=&\varepsilon_{ijk}A^{j}_{\mu}\dot{X}_{\mu}{\bf
I}_{k} \label{eomu2}
\end{eqnarray}
(\ref{eomu1}) and (\ref{eomu2}) are exactly equivalent to each
other. Their equivalence  is established by the following mapping
between the iso-spin $SU(2)$ generators and $S^2$ coordinates
 \begin{eqnarray}
 \frac{n_i}{2r}=\bar{u}{\bf I}_i u
 \end{eqnarray}
By this mapping, we can also show the following important identity
\begin{eqnarray}
\bar{u}[ {\bf A}_{\mu}, {\bf A}_{\nu}]u = -2ir^4F_{\mu
i}F_{ik}F_{k\nu} \label{AA}
\end{eqnarray}
Note that the $SU(2)$ field strength on $S^{4}$ is finite
everywhere while the $U(1)$ field strength is singular. For ${\bf
F}_{ab}$, it turns out that the singularity in $\partial_a {\bf
A}_b - \partial_b {\bf A}_a$ exactly cancels the singularity in
the commutator $[{\bf A}_a, {\bf A}_b]$. Therefore, the $SU(2)$
field strength ${\bf F}_{ab}$ and the U(1) $F_{\lambda\tau}$ have
the following correspondence
\begin{eqnarray} \nonumber
\bar{u}{\bf F}_{a5}u &=& -iF_{a5} \\
\bar{u}{\bf F}_{\mu\nu}{u} &=& -iF_{\mu\nu}-2ir^4F_{\mu
i}F_{ik}F_{k\nu}
\end{eqnarray}
Using these identities, and sandwiching the matrix equations
(\ref{eomu2}) by $\bar u$ and $u$ on both sides, we obtain
(\ref{eomu1}) for the case when $\partial V/\partial n=0$. This
proves the exact equivalence between the $CP_3$ equations of
motion obtained here and the $S^4$ non-abelian equations of motion
obtained in \cite{zhang2001}. The deep meaning of these remarkable
identities will become clear after we fuzzify $S^2$ in the
$CP_3\sim S^4\times S^2$ local decomposition.

\section{Semiclassical quantization and the flat space limit}
Similar to the 2D case, we can carry out a semi-classical
quantization of the action (\ref{single}).  In the 2D case, we
start with the Lagrangian (\ref{2dlagrangian}).  The action for
one orbit is given by
\begin{eqnarray}
S=\oint_{\lambda} A_{i}dx^{i} \label{closed}
\end{eqnarray}
where $\lambda$ is the closed contour on $S^{2}$.  The gauge
potential in (\ref{closed}) unlike the field strength $F$ suffers
from a Dirac string. Therefore, using Stoke's theorem, we can
convert the line integral to an area integral
\begin{eqnarray}
S=\frac{1}{2}\int_{u} F_{ij} dx^{i}\wedge dx^{j}
\end{eqnarray}
where $u$ denotes the upper hemisphere which is bounded by the
closed loop.  On the other hand, we can also integrate $F$ over
the lower hemisphere.  The difference between the two integrals is
the integral of $F$ over the whole of $S^2$.  This integral gives
$4\pi I$.  Since the two prescriptions should be equivalent, we
must set $4\pi I =2\pi n$ as a quantization condition.  This is
the Bohr-Sommerfeld quantization.  Thus, we obtain the
quantization $2I=n$.  $I$ is an integer or an half integer.

In our $CP_{3}$ case, the closed loop should be carefully chosen.
Since $H_{2}(CP_{3})=\textbf{Z}$, we can choose the closed loop on
a two-cycle $CP_{1}$.  $H_{2}(CP_{3})$ is the second homology
group of $CP_{3}$.  Define the map $\Xi:CP_3\rightarrow CP_{1}$ so
that the K\"ahler form on $CP_{3}$ maps to the one on $CP_{1}$.
We can always do that because $CP_{1}$ can be naturally embedded
into $CP_{3}$. Under this map, the loop integral of (\ref{single})
for one orbit becomes
\begin{eqnarray}
S = \oint_{\lambda} A_{\tau}dx^{\tau}=\oint_{\lambda} A_{i}dx^{i}
\label{semiquantization}
\end{eqnarray}
where $\lambda$ is a closed loop on $CP_1$\footnote{We use the
same notations here as in section $2$.  $\tau$ is in the set
\{$a$,$i$\}} and the $A_{i}'s$ are given by the second equation in
(\ref{connection}). The $A_{i}'s$ have the same form as those in
2D case.  Therefore, we obtain exactly the same quantization
condition as that in 2D.  $I$ is quantized to be an integer or an
half integer.

Just like the case of (\ref{2dlagrangian}), the single particle
Lagrangian can also be obtained by representing $CP_{3}$ as a
coset space $SU(4)/U(3)$.  The Lagrangian has a form similar to
(\ref{2dlagrangian})
\begin{eqnarray}
L=2iITr(YS^{-1}(x)\dot{S}(x)) \label{6dlagrangian}
\end{eqnarray}
Here $S(x)$ is a general group element of $SU(4)$.  In order to
mod by the stability group $U(3)$, we choose
\begin{eqnarray}
Y \!=\! \frac{1}{4}\left( \begin{array}{cccc}
               1  & 0 & 0 & 0  \\
               0  & 1 & 0 & 0 \\
               0  & 0 & 1 & 0 \\
               0  & 0 & 0 & -3\end{array} \right) \  \\
 \nonumber \label{SU4hypercharge}
\end{eqnarray}
Similar to the $2D$ case, it can be shown that
(\ref{6dlagrangian}) is equivalent to (\ref{splagrangian}).

This is a dynamical system with constraints. In order to quantize
(\ref{6dlagrangian}), we follow the Dirac constraint
formalism\cite{Diracbook} and the methods of \cite{balabook,Ydri}.
Let $T_{k}, k=1,...,15$, be the generators of $SU(4)$, satisfying
$Tr(T_{k}T_{k'})=\frac{1}{2}\delta_{kk'}$. We may choose $T_k$,
$k=1,...,8$, to be the generators of a $SU(3)$ subgroup. The
hypercharge of $SU(4)$ is chosen to be $Y=
\frac{3}{\sqrt{6}}T_{15}$. A general $SU(4)$ group element
$S(\xi)$ takes the form $S(\xi)= e^{iT_k \xi_k}$. $\xi_k$ is used
to parameterize the group manifold.  The momentum conjugate to
$\xi_k$ is given by $ \pi_k=\frac{\partial L}{\partial
\dot{\xi}_k}$. Let's define $S^{-1}dS=-iT_{k'}E_{k'k}d\xi_k$,
where $E$ is a $15 \times 15$ matrix and each element in $E$ is a
function of $\xi_k$. $\pi_k$ can be expressed as
\begin{eqnarray}
\pi_k = 2I Tr(YT_{k'}E_{k'k}) \label{pi}
\end{eqnarray}
Since $E_{k'k}$ is non-singular \cite{balabook}, we can introduce
 $\Lambda_k=-\pi_{k'}(E^{-1})_{k'k}$. (\ref{pi}) is simplified to
\begin{eqnarray}
\Lambda_k = -2I Tr(YT_{k})=-\frac{\sqrt{6}I}{2}\delta_{k,15}
\label{constraint1}
\end{eqnarray}
Using the Poisson brackets
\begin{eqnarray}
\{\xi_k,\xi_{k'}\}=\{\pi_k,\pi_{k'}\}=0, ~~
\{\xi_k,\pi_{k'}\}=\delta_{kk'}
\end{eqnarray}
one can easily show that $\Lambda_k$'s are the generators of
$SU(4)$ which act on the right of $S(\xi)$, i.e.
\begin{eqnarray}
\{\Lambda_k,S\}=-iST_k,
~~\{\Lambda_k,\Lambda_{k'}\}=f_{kk'k^{''}}\Lambda_{k^{''}}
\end{eqnarray}
where $f_{kk'k^{''}}$ are the structure constants of $SU(4)$.
Moreover, since the Lagrangian is of first order,
(\ref{constraint1}) essentially provides a set of constraints
determining the momentum conjugates.  They are expressed as
\begin{eqnarray}
\Lambda_k+\frac{\sqrt{6}I}{2}\delta_{k,15}\approx 0 \label{con}
\end{eqnarray}
The "$\approx$" means that the equalities are satisfied only on
the constraint surface.  For $k=1,...,8,15$ in (\ref{con}), the
constraints are first class constraints since the Poisson brackets
$\{\Lambda_k,\Lambda_{k'}\}$ weakly vanish on the constraint
surface. The rest of the constraints are the second class
constraints.  We can rearrange the second class constraints to
form a complete set of first class constraints. For $2I \geq 0$,
the set of new first class constraints is
\begin{eqnarray}
\Phi_{15}&=&\Lambda_{15}+\frac{\sqrt{6}I}{2}\approx0,~~
\Phi_{k}=\Lambda_k\approx 0, ~~k=1,...,8 \label{1st_class_constraint} \\
\Phi^-_{k}& = & \Lambda_k-i\Lambda_{k+1}\approx 0, ~~k=9,11,13
\label{2nd_class_constraint}
\end{eqnarray}
For $2I<0$, we rearrange the second class constraints by
$\Phi^+_{k}= \Lambda_k+i\Lambda_{k+1}\approx 0, ~~k=9,11,13$.  The
two sets of constraints map to different irreducible
representations of $SU(4)$.

Following the analysis in \cite{Ydri}, we can apply these
constraints on functions of $SU(4)$.  The states satisfying the
above constraints are $SU(3)$ singlets.  They have the
non-vanishing eigenvalue $-\frac{\sqrt{6}I}{2}$ of $\Lambda_{15}$
which acts on the right. Therefore, $2I$ must be an
integer\footnote{In other words, the states are the eigenstates of
Y with eigenvalues $-\frac{3}{4}(2I)$.}. The irreducible
representations (irreps) of $SU(4)$ are labelled by
$(n_1,n_2,n_3)$. For fixed $2I$, (\ref{1st_class_constraint}) and
(\ref{2nd_class_constraint}) select certain irreps.  They uniquely
determine $(2I,0,0)$ for $2I>0$, and $(-2I,-2I,-2I)=(-2I,0,0)^{*}$
for $2I<0$.   These representations consist of $SU(3)$ singlet
states which can not be raised or lowered any further by
(\ref{2nd_class_constraint}). This can be understood more easily
by looking at the weight diagram of the representations.  For
example, for $2I>0$, the conditions select the state at the bottom
in the weight diagram of $(2I,0,0)$. The constraint
(\ref{2nd_class_constraint}) serves the role of $SU(2)$ raising or
lowering operators. From the dynamical point of view, $CP_{3}$ is
the homogeneous space $SU(4)/U(3)$. We can view
(\ref{1st_class_constraint}) as the conditions which define the
coset space and (\ref{2nd_class_constraint}) as the force normal
to the constraint surface.  The force being zero on the surface
indicates that particles only move along $CP_3$.  This completes
the quantization procedure.  $(2I,0,0)$ and $(2I,0,0)^*$ are the
symmetrical tensor irreps of $SU(4)$ with the dimension equal to
$\frac{1}{6}(|2I|+1)(|2I|+2)(|2I|+3)$. They are also identical to
the spinor irreps $(|2I|,0)$ of $SO(5)$. These irreps are exactly
the lowest-Landau-level functions
\begin{eqnarray}
\sum_{m_1+m_2+m_3+m_4=2I} \sqrt{\frac{(2I)!}{m_1! m_2! m_3! m_4!}}
\Psi_1^{m_1} \Psi_2^{m_2} \Psi_3^{m_3} \Psi_4^{m_4}
\end{eqnarray}
found in equation (9) of ref. \cite{zhang2001}. The
lowest-Landau-level states of the $SO(5)$ symmetric Hamiltonian
defined in ref. \cite{zhang2001} indeed have a larger, $SU(4)$
symmetry group. States including higher Landau levels have only
$SO(5)$ symmetry.

The physical picture of this $CP_3$ model can be visualized more
clearly by taking the flat space limit.  Take (\ref{single}), and
expand it around $X_{5}=R$ and $n_{3}=r$.  It becomes
\begin{eqnarray}
L=-I\varepsilon_{3ij}\frac{n_{i}\dot{n_{j}}}{2r^{2}}-I\varepsilon_{3ij}\frac{X_{i}\dot{X_{j}}}{2R^{2}}-I\varepsilon_{12\mu\nu}\frac{X_{\mu}\dot{X_{\nu}}}{2R^{2}}
\end{eqnarray}
In this limit, the system behaves like three independent 2D QHE.
 The three independent planes are $(n_{1},n_{2})$, $(X_{1},X_{2})$,
and $(X_{3},X_{4})$. The noncommutative algebra is as what
follows:
\begin{eqnarray}
[\frac{n_{1}}{2r}, \ \frac{n_{2}}{2r}]=\frac{i}{2I} \label{n-ncg}
\end{eqnarray}
\begin{eqnarray}
[X_{1}, \ X_{2}]=\frac{2iR^{2}}{I}
\end{eqnarray}
\begin{eqnarray}
[X_{3}, \ X_{4}]=\frac{2iR^{2}}{I}
\end{eqnarray}
In $(\ref{n-ncg})$, $\frac{n_{i}}{2r}$ plays the role of a
classical spin, and its commutation relation vanishes as the
classical limit $I\rightarrow\infty$. These quantization
equations agree exactly with the non-commutative geometry
equation (14) of ref. \cite{zhang2001}, when it is expanded
around $n_3=r$.

\section{Field theory Lagrangian}
We go from the single particle Lagrangian (\ref{single}) to the
Lagrangian of many particles
\begin{eqnarray}
\int dt \sum_i A_B(X_i) \partial_t X_i^B = \int dt d^6 x \rho(x)
A_B \partial_t X^B =\int dt d^6 x A_B J^B \label{many}
\end{eqnarray}
where $\rho(x)=\sum_i \delta(x-X^i)$ is the particle density of
the liquid, and $J^A=\rho\partial_t X^A$ is the particle current
density. They satisfy the equation of continuity
\begin{eqnarray}
\partial_\Gamma J^\Gamma = \partial_t \rho + \partial_A J^A =0
\label{continuity}
\end{eqnarray}

The key to the Chern-Simons construction in the $2+1$ dimensional
QHE is the flux attachment
transformation\cite{zhang1989,zhang1992}. The particle is attached
to its dual, the $2$-form field strength
$J^\mu=\nu\epsilon^{\mu\nu\rho}\partial_\nu A_\rho$. A conserved
current in $6+1$ dimensions is equivalent to a $6$-form field
strength with the continuity equation (\ref{continuity}) replaced
by the Bianchi identity of the $6$-form field strength. Thus the
natural generalization in the higher dimensional case is to attach
to the particle its dual, the six form field strength, {\it i.e.}
\begin{eqnarray}
\rho = {\nu \over 8 \ 3! } \epsilon^{ABCDEG} {\cal F}_{AB} {\cal
F}_{CD} {\cal F}_{EG} \label{rho}
\end{eqnarray}
From the equation of continuity we also obtain the generalization:
\begin{eqnarray}
J^\Gamma = {\nu \over
3!}\epsilon^{\Gamma\Gamma_1\Gamma_2\Gamma_3\Gamma_4\Gamma_5\Gamma_6}
(\partial_{\Gamma_1} {\cal A}_{\Gamma_2}) (\partial_{\Gamma_3}
{\cal A}_{\Gamma_4}) (\partial_{\Gamma_5} {\cal
A}_{\Gamma_6})\label{current}
\end{eqnarray}
From this representation we see that the equation of continuity
(\ref{continuity}) is automatically satisfied. The gauge potential
and the field strength introduced here have two components, ${\cal
A}=A+a$ and ${\cal F}=F+f$, where $A$ and $F$ are the background
$U(1)$ gauge potential and field strength defined on the $CP_3$
manifold, which are explicitly given in equations
(\ref{connection}) and (\ref{F}). $a$ and $f$ are the dynamically
fluctuating parts of the gauge potential and the field strength;
they describe deviations from the equilibrium density and current.
In the equilibrium ground state, $a=f=0$, and from equation
(\ref{rho}) we see that the uniform ground state particle density
is proportional to the background flux density, with the constant
of proportionality being the filling factor $\nu$:
\begin{eqnarray}
\bar\rho = {\nu \over 8\ 3! } \epsilon^{ABCDEG} {F}_{AB} {F}_{CD}
{F}_{EG} \label{rhobar}
\end{eqnarray}
It can be shown that $F$ is also the K\"ahler form of the $CP_3$
manifold \cite{nair}\footnote{Also see the Appendix.}, and $F
\wedge F \wedge F$ is nothing but the volume form, which is
uniform over the $CP_3$ manifold. Integrating both sides of the
equation (\ref{rhobar}) we see that $N\sim R^6$, which agrees with
the scaling obtained by ZH\cite{zhang2001}.

The generalized flux attachment equation (\ref{current}) can be
naturally obtained from the functional variation with respect to
the following CS action:
\begin{eqnarray}
\int dt d^6x ({\nu \over 4 \ 3!}
\epsilon^{\Gamma\Gamma_1\Gamma_2\Gamma_3\Gamma_4\Gamma_5\Gamma_6}
{\cal A}_{\Gamma} (\partial_{\Gamma_1} {\cal A}_{\Gamma_2})
(\partial_{\Gamma_3} {\cal A}_{\Gamma_4}) (\partial_{\Gamma_5}
{\cal A}_{\Gamma_6}) - J^\Gamma {\cal A}_\Gamma) \label{cs}
\end{eqnarray}
The coefficient of the CS term is given by $\nu$, identified with
the filling factor of the quantum Hall fluid.

There is a precise way in which the fluid dynamics can be
described by maps of $CP_3$ onto itself. To establish this
connection, let us work in a frame ``comoving'' with the particles
in which the velocity fields are taken to be zero and the density
fixed in time, and specialize to the ${\cal A}_0=0$ gauge. We have
previously introduced the $\Psi$ spinor as the coordinate of a
single particle on the $CP_3$ manifold. In order to describe the
collection of particles, or a continuous fluid, we can promote
$\Psi$ into a $CP_3$ non-linear sigma model field $\Psi(x,t)$ over
the base $CP_3$ manifold with local coordinate $x$. The
four-component complex spinor field $\Psi_{\alpha}(x)$ is subject
to the following identification
\begin{equation}
\bar{\Psi}\Psi=1, \,\ \{\Psi_\alpha\}\sim
e^{i\theta}\{\Psi_\alpha\}\label{emb}
\end{equation}
It is easy to see that when the gauge function $\theta$ is taken
to be a local function on $CP_3$, the field
\begin{equation}
{\cal A}_{\Gamma}=2iI\bar{\Psi}\partial_{\Gamma}\Psi
\label{transf}
\end{equation}
transforms as a $1$-form gauge potential. We identify this field
with the gauge field appearing in the action (\ref{cs}). For later
convenience, we may write the above expression in terms of
differential forms: ${\cal A}=iI\bar{\Psi}d\Psi$. The gauge
invariant field strength is given by
\begin{equation}
{\cal F}=2iI d\bar{\Psi}\wedge d\Psi
\end{equation}
Substituting this into the expression for the density $\rho$, we
obtain
\begin{equation}
\rho={\nu I^3 \over 3!} (2i d\bar{\Psi}\wedge d\Psi)\wedge
(2id\bar{\Psi}\wedge d\Psi) \wedge (2i d\bar{\Psi}\wedge d\Psi)
\label{rho2}
\end{equation}
Hence, $\int \rho$ exactly measures the winding number of the
$CP_3\rightarrow CP_3$ mapping. On the other hand, if we
substitute (\ref{transf}) into the expression for the Chern-Simons
action and make use of the identifications in ($\ref{emb}$), we
obtain within the ${\cal A}_0=0$ gauge
\begin{equation}
S=2iI\int dtd^6x \rho(x)\bar{\Psi}\partial_t\Psi \label{many1}
\end{equation}
This agrees exactly with the many particle fluid action
(\ref{many}). The action is invariant under the gauge
transformation appearing in (\ref{emb}).

The gauge transformations of the $\Psi$'s are naturally thought of
as volume preserving, time independent coordinate transformations
under which the expression for the density (\ref{rho2}) remains
invariant. Since $CP_3$ is a K\"ahler manifold the volume form is
given by $dV=( F \wedge F \wedge F)/ 3!$ where $F$ is the $2$-form
K\"ahler metric. Therefore, coordinate transformations that
preserve the K\"ahler form are also volume preserving. These are
euivalent to the $U(1)$ gauge transformations in (\ref{emb}). In
this picture, the $CP_3$ manifold of the incompressible quantum
liquid can be viewed as a phase space. Liouville's theorem
requires that the volume of phase space is conserved and so volume
preserving transformations of the $\Psi$'s are gauged.

{\bf{Quasiparticles}}. Maps of $CP_3$ onto itself are
characterized by the topological invariant (or winding number)
\begin{eqnarray}
Q_F=\frac{1}{3!} \int_{CP_3} {\cal F} \wedge {\cal F} \wedge {\cal
F} = \frac{8I^3}{3!} \int_{CP_3} (id\bar{\Psi}\wedge d\Psi) \wedge
(id\bar{\Psi}\wedge d\Psi) \wedge (id\bar{\Psi}\wedge d\Psi)
\end{eqnarray}
$Q_F$ scales with $I^3 \propto \Omega_{CP_3}$, where
$\Omega_{CP_3}$ is the volume of the $CP_3$ manifold.  From
equation (\ref{rho}) we see that the total charge of the fluid is
$Q=\nu Q_F$. Since $Q_F$ is a topological winding number, it can
only change by an integer value. A quasi-particle or a quasi-hole
is created by $\Delta Q_F=\pm 1$. In this case, equation
(\ref{rho}) implies that the charge of the quasi-particle or
quasi-hole is given by
\begin{equation}
\Delta Q= \pm \nu .
\end{equation}
which confirms the value of the fractional charged obtained
in\cite{zhang2001}. As in Laughlin's theory, quasi-particles or
quasi-holes have one unit of magnetic flux attached to them. The
effective field theory does not predict the allowed value of the
fractional filling factor, only the microscopic theory restricts
$\nu=1/m^3$\cite{zhang2001}.

\section{Dimensional reduction and extended objects}
The $CP_3\rightarrow CP_3$ mappings describe point-like
topological excitations. Condensed matter systems also support
elementary excitations which are extended objects. The dimension
of the extended objects is determined by the homotopy class of the
base and the order parameter spaces \cite{Mermin}. For example, a
superfluid with $U(1)$ order parameter in $D$ spatial dimensions
supports $D-2$ dimensional extended objects, or $(D-2)$ branes in
modern string theory language. We show here that the QH liquid
constructed by ZH also supports $2$-brane (membrane) and $4$-brane
excitations, which are associated with the maps of
$CP_3\rightarrow S^4$ and $CP_3\rightarrow S^2$. In this section
we will describe these objects, closely following the field
theoretical treatment of Wu and Zee \cite{Wu+Zee} and that of
Wilczek and Zee \cite{Wilczek+Zee}.

Thick solitonic membranes can be constructed from $CP_3$
$\rightarrow$ $S^4$ mappings. They wrap a spherical $2$-cycle of
$CP_3$ and are characterized by the topological current
\begin{equation} J^{\Gamma\Gamma_1\Gamma_2}={1 \over 4!}
\epsilon^{\Gamma\Gamma_1\Gamma_2\Gamma_3\Gamma_4\Gamma_5\Gamma_6}
\epsilon_{abcde} X^a\partial_{\Gamma_3} X^b \partial_{\Gamma_4}
X^c
\partial_{\Gamma_5} X^c
\partial_{\Gamma_6} X^e
\label{J_3form} \end{equation} where the $X^a(x)$'s are $O(5)$
sigma model-like fields constructed from the embedding fields
$\Psi(x)$ by using the second Hopf map: $X^a(x)/R = \bar{\Psi}(x)
\Gamma^a \Psi(x)$. The membranes can be thought of as non-trivial
topological configurations build out of $O(5)$ sigma-model fields
that wind around a spherical $4$-cycle in $CP_3$. The
non-vanishing of the homotopy group $\pi_4(S^4) =Z$ ensures the
stability of such configurations. The conserved topological
charges are given by
\begin{equation} Q^{\Gamma_1 \Gamma_2} = \frac{1}{\Omega_4} \int_{S^4} J^{0
\Gamma_1 \Gamma_2} \end{equation} \noindent There are $15$
independent such charges, all integers, and their combinations can
be used to classify the maps.

Similarly, we can build thick solitonic $4$-branes in our fluid.
These ojects wrap spherical $4$-cycles of $CP_3$. They are
constructed as maps $CP_3 \rightarrow S^2$, and they are
characterized by the topological current
\begin{eqnarray}
J^{\Gamma\Gamma_1\Gamma_2\Gamma_3\Gamma_4} ={1 \over 2}
\epsilon^{\Gamma\Gamma_1\Gamma_2\Gamma_3\Gamma_4\Gamma_5\Gamma_6}
\epsilon_{ijk} n^i\partial_{\Gamma_5} n^j \partial_{\Gamma_6} n^k
\label{J_5form}
\end{eqnarray}
\noindent Here, the $n^i(x)$'s are $O(3)$ sigma model-like fields
given by $n^i(x)/r = \bar{u}(x) \sigma^i u(x)$, with $u$ given in
Eq. (\ref{Psi}) and (\ref{u}). The $4$-branes are non-trivial
topological configurations build out of $O(3)$ sigma model fields
winding a spherical $2$-cycle of $CP_3$. The stability of the
$4$-branes lies in the non-vanishing of the homotopy group
$\pi_2(S^2)=Z$. The conserved topological charges are given by
\begin{eqnarray} Q^{\Gamma_1\Gamma_2\Gamma_3\Gamma_4} =\frac{1}{\Omega_2}
\int_{S^2} J^{0\Gamma_1\Gamma_2\Gamma_3\Gamma_4} \end{eqnarray}
\noindent As in the membrane case there are $15$ such charges, all
integers, and their combinations can be used to classify the maps.

The membranes and $4$-branes described above couple to $p$ ($p$=3
and $p$=5) form gauge fields, through the Lagrangians
\begin{eqnarray}
\int d^7 x C_{\Gamma\Gamma_1\Gamma_2} J^{\Gamma\Gamma_1\Gamma_2}
\end{eqnarray}
\noindent and
\begin{eqnarray}
\int d^7 x C_{\Gamma\Gamma_1\Gamma_2\Gamma_3\Gamma_4}
J^{\Gamma\Gamma_1\Gamma_2\Gamma_3\Gamma_4}
\end{eqnarray}
\noindent These Lagrangians are analogs of (\ref{many}); they can
describe a finite density of extended objects.

We may consider the membranes and $4$-branes in the thin-brane
limit and write the currents as \begin{equation}
J^{\Gamma\Gamma_1\Gamma_2}(y) = \int d\sigma_0 d \sigma_1  d
\sigma_2 \delta^7 (y-X(\sigma_0,\sigma_1,\sigma_2))
\frac{\partial({X^{\Gamma}, X^{\Gamma_1}, X^{\Gamma_2}})}{\partial
({\sigma_0, \sigma_1, \sigma_2})}
\end{equation} \noindent and \begin{equation} J^{\Gamma\Gamma_1 \Gamma_2 \Gamma_3 \Gamma_4}(y)
= \int d\sigma_0 d \sigma_1 d \sigma_2 d \sigma_3 d \sigma_4
\delta^7 (y-X(\sigma_0,\sigma_1,\sigma_2, \sigma_3, \sigma_4))
\frac{\partial({X^{\Gamma}, X^{\Gamma_1}, X^{\Gamma_2},
X^{\Gamma_3}, X^{\Gamma_4}})}{\partial({\sigma_0, \sigma_1,
\sigma_2, \sigma_3, \sigma_4})} \end{equation} where $(\sigma_0,
\sigma_i)$ are the world-volume coordinates of the extended object
and $\partial(\ldots) /
\partial(\ldots)$ denotes a Jacobian.

In the limit of thin objects, the ``single p-brane" Lagrangians
are a generalization of the single particle Lagrangian
(\ref{single}), and are given by \cite{Lund+Regge} and
\cite{Matsuo+Shibusa}
\begin{eqnarray} \label{lagrangian1}
\int d\sigma_0 d\sigma_1 d\sigma_2 C_{\Gamma\Gamma_1\Gamma_2}(X)
\epsilon^{\sigma_0\sigma_1\sigma_2}
\partial_{\sigma_0} X^{\Gamma} \partial_{\sigma_1} X^{\Gamma_1}
\partial_{\sigma_2} X^{\Gamma_2}
\end{eqnarray}
and
\begin{eqnarray} \label{lagrangian2}
\int d\sigma_0 d\sigma_1 d\sigma_2 d\sigma_3 d\sigma_4
C_{\Gamma\Gamma_1\Gamma_2\Gamma_3\Gamma_4} (X)
\epsilon^{\sigma_0\sigma_1\sigma_2\sigma_3\sigma_4}
\partial_{\sigma_0} X^{\Gamma} \partial_{\sigma_1} X^{\Gamma_1}
\partial_{\sigma_2} X^{\Gamma_2} \partial_{\sigma_3} X^{\Gamma_3}
\partial_{\sigma_4} X^{\Gamma_4}
\end{eqnarray}
where $(\sigma_0,\sigma_i)$ are the world-volume coordinates of
the extended object. The higher form gauge fields felt by the
extended objects can be constructed explicitly out of the $1$-form
background gauge field $A$ \cite{Wu+Zee}, explicitly given in
(\ref{connection})
\begin{eqnarray}
C_{3}= A \wedge dA, \,\ C_{5}={1 \over 2}A \wedge dA \wedge dA
\end{eqnarray}
From Eq.(\ref{lagrangian1}) (\ref{lagrangian2}) we can obtain the
equations of motion for membranes and $4$-branes in the background
$p$ form gauge field.

We start with the membrane. The single membrane Lagrangian in
Eq.(\ref{lagrangian1}) can be cast in the form
\begin{displaymath}
\int d\sigma_0 d\sigma_1 d\sigma_2 \epsilon^{\sigma_0 \sigma_1
\sigma_2}
\partial_{\sigma_0} X^{\Gamma} \partial_{\sigma_1} X^{\Gamma_1} \partial_{\sigma_2}
X^{\Gamma_2} C_{\Gamma \Gamma_1 \Gamma_2}(X)
\end{displaymath}
\begin{displaymath}
= \int d\sigma_0 d\sigma_1 d\sigma_2 \epsilon^{\sigma_0 \sigma_1
\sigma_2} X^{\Gamma}
\partial_{\sigma_1} X^{\Gamma_1}
\partial_{\sigma_2} X^{\Gamma_2} [\partial_{\Gamma_3} C_{\Gamma
\Gamma_1 \Gamma_2}(X)] \partial_{\sigma_0} X^{\Gamma_3}
\end{displaymath}
\begin{equation} \label{singlemembranecast}
= -\frac{1}{4}\int d\sigma_0 d\sigma_1 d\sigma_2
\epsilon^{\sigma_0 \sigma_1 \sigma_2} F_{\Gamma \Gamma_1 \Gamma_2
\Gamma_3} X^{\Gamma}
\partial_{\sigma_0} X^{\Gamma_1}
\partial_{\sigma_1} X^{\Gamma_2}  \partial_{\sigma_2} X^{\Gamma_3}
\end{equation}
\noindent where we have defined the field strength of the $3$-form
$C$, $F=dC$ by
\begin{equation} \label{3formfieldstrength}
F_{\Gamma \Gamma_1 \Gamma_2 \Gamma_3} = \partial_{\Gamma}
C_{\Gamma_1 \Gamma_2 \Gamma_3} +\partial_{\Gamma_1} C_{\Gamma_3
\Gamma_2 \Gamma} +\partial_{\Gamma_2} C_{\Gamma_3 \Gamma \Gamma_1}
+\partial_{\Gamma_3} C_{\Gamma_1 \Gamma \Gamma_2}
\end{equation}
\noindent We can therefore now write the Lagrangian density in the
2+1 dimensional $(\sigma_0, \sigma_i)$ world volume as
\begin{equation}
{\cal{L}} ={-\frac{1}{4}}\epsilon^{\sigma_0 \sigma_1 \sigma_2}
F_{\Gamma \Gamma_1 \Gamma_2 \Gamma_3} X^{\Gamma}
\partial_{\sigma_0} X^{\Gamma_1}
\partial_{\sigma_1} X^{\Gamma_2}  \partial_{\sigma_2} X^{\Gamma_3}
\end{equation}
\noindent And therefore the equation of motion for the membrane in
the presence of the $3$-form gauge field $C$ now becomes
\begin{equation} \label{membranequationofmotion1}
F_{\Gamma \Gamma_1 \Gamma_2 \Gamma_3} \epsilon^{\sigma_0 \sigma_1
\sigma_2}
\partial_{\sigma_0} X^{\Gamma_1}
\partial_{\sigma_1} X^{\Gamma_2}  \partial_{\sigma_2} X^{\Gamma_3}
= 0
\end{equation}
\noindent Going through exactly the same steps, one can obtain the
equation of motion for the $4$-brane Lagrangian
Eq.(\ref{lagrangian2}). This reads
\begin{equation} \label{membranequationofmotion2}
F_{\Gamma \Gamma_1 \Gamma_2 \Gamma_3 \Gamma_4 \Gamma_5}
\epsilon^{\sigma_0 \sigma_1 \sigma_2 \sigma_3 \sigma_4}
\partial_{\sigma_0} X^{\Gamma_1}
\partial_{\sigma_1} X^{\Gamma_2}  \partial_{\sigma_2} X^{\Gamma_3}
\partial_{\sigma_3} X^{\Gamma_4} \partial_{\sigma_4} X^{\Gamma_5}= 0
\end{equation} \noindent where $F_{\Gamma \Gamma_1 \Gamma_2 \Gamma_3
\Gamma_4 \Gamma_5}$ is the field strength associated with the
$5$-form $C$.

A particularly nice solution of the equation of motion is that of
a membrane extending in the $X^{5}$ and $X^{6}$ field directions
and moving along the other directions perpendicular to the
magnetic field. The solution has the form
\begin{equation}
X^0 =\sigma_0, \ X^{a} =f_a(\sigma_0), \ a=1,...,4 \label{solu1}
\end{equation} \noindent
while at the same time
\begin{equation}
X^{5} =f(\sigma_1, \sigma_2),\ X^{6} =g(\sigma_1, \sigma_2)
\label{solu2}
\end{equation} \noindent
Extended membrane solutions stand if the Jacobian $
\partial(f,g)/\partial(\sigma_{1},\sigma_{2})$ is not zero. If it vanishes, we have a
particle solution on $CP_{3}$. Plugging (\ref{solu1}) and
(\ref{solu2}) into the equation of motion
(\ref{membranequationofmotion1}), we have
\begin{eqnarray}
F_{\Gamma\Gamma_1 56} \epsilon^{\sigma_0 \sigma_1 \sigma_2}
\partial_{\sigma_0} X^{\Gamma_1}
\partial_{\sigma_1} f  \partial_{\sigma_2} g
= 0
\end{eqnarray}
This simply says that the membrane has to move in the $X^a$
directions perpendicular to the effective magnetic field:
$B_{\Gamma\Gamma_1}=F_{\Gamma\Gamma_1 56}$.

From this solution we can see that the membrane behaves
effectively like a particle in the $X^a$ space. The dynamical
degrees of freedom of the membrane is described by a $O(3)$
non-linear $\sigma$ model on the world volume of the membrane.
These degrees of freedom can be discretized by ``fuzzifying" the
$S^2$ sphere, in which case they appear as iso-spin quantum
numbers of the particle.

\section{Dimensional reduction using fuzzy spheres}
So far, we have treated the extended objects as continuous
objects. However, it is also possible to treat them as a
point-like objects with a large number of internal degrees of
freedom. This can be done by fuzzifying the $S^2$ or $S^4$ in the
$CP_3=S^4\times S^2$ decomposition.

For the reduction on a fuzzy $S^2$, the Cartesian coordinates
$n_i$ are replaced by $SU(2)$ matrices ${\bf{I}}_i$, and the
continuous integral over $S^2$ is replaced by the trace over the
$SU(2)$ matrices. Under this procedure, the $U(1)$ Chern-Simons
theory over $CP_3$ is dimensionally reduced to a non-abelian
$SU(2)$ Chern-Simons theory over $S^4$.

We first briefly review the non-abelian theory. The Lagrange
density is given by\footnote{In this section, we use conventions
so that the field theory gauge field {\bf A} is Hermitian. Its
background expectation value is given by $i{\bf A}_{back}$, where
${\bf A}_{back}$ is given explicitly in Eq. (\ref{su2strength}).
The field strength is given by ${\bf
F}_{\mu\nu}=\partial_{\mu}{\bf A}_{\nu}-\partial_{\nu}{\bf
A}_{\mu}-i[{\bf A}_{\mu} - {\bf A}_{\nu}]$.}
\begin{equation} {\cal{L}}=\mu
Tr\left({\bf A} \wedge d{{\bf A}} \wedge d{\bf A} - {3i \over
2}{\bf A} \wedge {\bf A} \wedge {\bf A} \wedge d{{\bf A}}- {3
\over 5}{\bf A} \wedge {\bf A} \wedge {\bf A} \wedge {\bf A}\wedge
{\bf A} \right) \label{nacs} \end{equation} Varying the action
with respect to ${\bf A}$, we obtain
\begin{equation} \delta
{\cal{L}}_{CS}= 3\mu Tr(\delta {\bf A} \wedge {\bf F} \wedge {\bf
F}) \end{equation} and so, as in the abelian case, the equation of
motion can be taken to be ${\bf F} \wedge {\bf F}=0$. We may also
add a ``background charge'' density term\footnote{The four form
{\bf J} is the Hodge dual of the gauge covariant current ${\bf
J}^{\mu}$.}
\begin{equation}
\delta{\cal{L}}=-Tr({\bf A}_0dt\wedge {\bf J})
\end{equation}
Then, in terms of the four-form ${\bf J}$, the ${\bf A}_0$
equation of motion is replaced by
\begin{equation}
{\bf F} \wedge {\bf F}={{\bf J} \over 3\mu} \label{eqnmna}
\end{equation}

We choose the Cartesian co-ordinates $n_i$, given by
(\ref{1sthopf}) and (\ref{u}) to parameterize the isospin sphere.
In the $CP_3=S^4 \times S^2$ decomposition, the dynamical gauge
invariant fields can be taken to be the two sets of sigma-model
fields: an $O(5)$ sigma model field $X_a(x)$ and an $O(3)$ sigma
model field $n_i(x)$. We consider the case where the $O(3)$
sigma-model fields get fixed expectation values
\begin{equation}
\frac{\langle n_i(x)\rangle}{r} = \bar{u}\sigma_i u
\end{equation}
and consider field configurations $X_a(x)$ that are homogeneous
over the internal isospin sphere. Expanding the fluid gauge field
introduced in (\ref{transf}), we obtain to the leading order in
the fluctuations
\begin{equation}
{{\cal A} \over
2I}=i\bar{\Psi}d\Psi=i\bar{u}du+\hat{A}=i\bar{u}du+\bar{u}\hat{A}^i{\sigma_i
\over 2}u=i\bar{u}du+ \hat{A}_a^i{n_i\over 2r}dx^a \label{expand}
\end{equation}
Allowing the fluctuating field $\hat{A}^i\sigma_i/2$ to transform
as an $SU(2)$ gauge field, ${\cal A}$ is invariant under local
$SU(2)$ rotations of the spinor $u$. Therefore, our theory enjoys
a large symmetry consisting of $SU(2)$ gauge transformations. The
field strength gets an expectation value
\begin{equation}
\langle {\cal F} \rangle=2iId\bar{u}\wedge du
\end{equation}
The last expression is the magnetic field of a $U(1)$ monopole at
the center of the $2$-sphere: $F_{ij}=I\epsilon_{ijk}n_k/r^3$.

We now substitute the expansion (\ref{expand}) into the
Chern-Simons action (\ref{cs}), and average over the internal
isospin sphere. To do this, we first promote the fluctuating gauge
field $\hat A(n)$ into a non-abelian gauge field as follows. We
replace the $O(3)$ sigma-model field with non-commuting isospin
operators
\begin{equation}
 {In_i \over 2r}\leftrightarrow {{\bf{{I}}}_i}, \,\ [{\bf{{I}}}_i,
\,\ {\bf{{I}}}_j]=i\epsilon_{ijk}{\bf{{I}}}_k
\end{equation}
Thus the fields become matrices and the isospin sphere is replaced
by a fuzzy two-sphere. Integrals over over the internal space can
be replaced by
\begin{equation}
{1 \over 2\pi}\int_{S^2} \langle {\cal F}\rangle \ldots
\leftrightarrow TrI_d \ldots \label{dimrep}\end{equation} In this
way, the dimension of the $SU(2)$ representation is determined by
the strength of the monopole charge: $d=2I+1$. Derivatives with
respect to the isospin co-ordinates $n_i$ can be replaced by
commutators
\begin{equation}
i\epsilon_{ijk}\partial_j \hat{V}{In_k} =ir^3F_{ij}\partial_j
\hat{V} \leftrightarrow [{\bf {I}}_i, \,\ {\bf V}]
\end{equation}
From this identification, we learn that the commutator between any
two representation matrices can be obtained from
\begin{equation}
2ir^4\partial_i\hat{V}_1F_{ij}\partial_j\hat{V}_2 \leftrightarrow
[{\bf V}_1, \,\ {\bf V}_2] \label{commutator}
\end{equation}
With these identifications the definition of the covariant
derivative and the field strength follows. Finally, we point out a
useful identity \begin{equation} dn_i \wedge dn_j =
2ir\epsilon_{ijk}n_k d\bar{u}\wedge du \label{id}
\end{equation}

Our matrices satisfy
\begin{equation}
Tr\ {\bf{I}}_i\ {\bf{I}}_j={d C_2 \over 3}\delta_{ij}\label{trace}
\end{equation}
where the numbers $d=2I+1,\,\ C_2=I(I+1)$ denote the dimension and
Casimir of the $SU(2)$ representation. For large $I$,
(\ref{trace}) agrees with the expression of $\int_{S^2}F (I^2 n_i
n_j)/2\pi r^2$.

The Lagrange density of the $6+1$ dimensional Chern-Simons theory
is given by
\begin{equation} {\cal{L}}={\nu \over 4 \ 3!} {\cal A} \wedge
d{\cal A} \wedge d{\cal A} \wedge d{\cal A} - {\cal A}_0 \rho
\label{lag6}
\end{equation}
For simplicity, we may work in the ${\cal A}_0=0$ gauge and impose
its equation of motion as a constraint
\begin{equation}
{\nu \over 3!}{\cal F}\wedge {\cal F}\wedge {\cal F}=\rho
\label{constraint}
\end{equation}
In this gauge, the symmetry of the problem reduces to the group of
time independent gauge transformations. Expanding the first term
in (\ref{lag6}) about $\langle F \rangle$, we get two types of
terms contributing \footnote{We carried an integration by parts}
\begin{equation}
{\nu I^3 \over 3}\left(4(iId\bar{u}\wedge du)\wedge \hat{A} \wedge
d\hat{A} \wedge d\hat{A}+ I\hat{A} \wedge d\hat{A} \wedge d\hat{A}
\wedge d\hat{A}\right)
\end{equation}
When written in the ${\cal A}_0=0$ gauge, this becomes
\begin{equation}
{\nu I^3 \over 3} \left[ 4\left(\hat{A} \wedge
(\partial_t{\hat{A}}\wedge dt) \wedge d\hat{A} + \hat{A} \wedge
d\hat{A}\wedge (\partial_t{\hat{A}}\wedge dt)\right) \wedge
(iId\bar{u} \wedge du)+ 3I\hat{A}\wedge (\partial_t{\hat{A}}
\wedge dt) \wedge d\hat{A} \wedge d\hat{A} \right]
\end{equation}
In the last term, since $\hat A \sim dx$, the operator $d$ has to
take derivatives with respect to the 2 dimensional isospin space,
in order to satisfy the total antisymmetry. Taking this into
account, the last term is given explicitly by
\begin{equation}
I\hat{A} \wedge (\partial_t{\hat{A}} \wedge dt) \wedge d_2\hat{A}
\wedge d_2\hat{A}={I \over 4r^2}\hat{A} \wedge
(\partial_t{\hat{A}} \wedge dt) \wedge (\hat{A}^i_{\mu}
\hat{A}^j_{\nu}dx^{\mu}\wedge dx^{\nu})\wedge dn_i \wedge dn_j
\end{equation}
Using (\ref{id}), we may simplify this as follows
\begin{equation}
{1 \over 2}\hat{A} \wedge (\partial_t{\hat{A}} \wedge dt) \wedge
(\epsilon_{ijk}\hat{A}^i_{\mu} \hat{A}^j_{\nu}{In_k \over r}
dx^{\mu}\wedge dx^{\nu}) \wedge (id\bar{u} \wedge du)
\end{equation}
Now, we may replace the term in second parenthesis with a
commutator using (\ref{commutator}). We end up with
\begin{equation}
{-2i}\hat{A} \wedge (\partial_t{\hat{A}} \wedge dt) \wedge ({\bf
A} \wedge {\bf A}) \wedge (id\bar{u} \wedge du)
\end{equation}

Averaging over the isospin sphere and making the replacements
mentioned in the previous paragraph gives \begin{equation}
\langle{\cal{L}}\rangle_{S^2}={4\pi\nu \over 3 }Tr\left( {\bf A}
\wedge (\partial_t{{\bf A}}\wedge dt) \wedge d{\bf A} +{\bf
A}\wedge d{\bf A} \wedge (\partial_t{{\bf A}}\wedge dt) - {3i\over
2}{\bf A} \wedge {\bf A} \wedge {\bf A} \wedge (\partial_t{{\bf
A}} \wedge dt)\right) \label{gauge}
\end{equation}
This is the Lagrange density obtained from the non-abelian
Chern-Simons Lagrangian (\ref{nacs}) in the temporal gauge ${\bf
A}_0=0$, with $\mu$ = $4\pi \nu/3$. In the temporal gauge, there
is a left over symmetry consisting of time independent,
infinitesimal $SU(2)$ gauge transformations. As we already
mentioned, this symmetry is a symmetry of the full theory. Because
of this symmetry, there exist three local conserved quantities,
one for each independent $SU(2)$ rotation. It is easily verified,
using for example the equations of motion by varying
(\ref{gauge}), that ${\bf F} \wedge {\bf F}$ is time independent.
Therefore, we may consistently set
\begin{equation}
{4\pi \nu }{\bf F} \wedge {\bf F}= {\bf J}
\end{equation}
with ${\bf J}$ time independent and treat this as a constraint
equation. We normalized the equation to be exactly the same as the
${\bf A}_0$ equation of motion (\ref{eqnmna}). Thus we may
introduce the components of ${\bf A}_0$ into the action as a
Lagrange multiplier and obtain the full Chern-Simons action
(\ref{nacs}) together with a background charge density term.

There is a constraint on $Tr{\bf J}$ from (\ref{constraint}).
Expanding the density $\rho$ about $\langle F\rangle$ gives
\begin{equation}
\rho=4\nu I^3  \left[(id\bar{u}\wedge du)\wedge d\hat{A}\wedge
d\hat{A}+d_4\hat{A}\wedge d_2\hat{A}\wedge d_2\hat{A}\right]
\end{equation}
The last term can be worked out as before to give
\begin{equation}
\rho=4\nu \left( I^2d\hat{A}\wedge d\hat{A}-2iI({\bf A}\wedge {\bf
A})\wedge d\hat{A}\right)\wedge(iId\bar{u}\wedge du)
\end{equation}
Averaging over the two-sphere, we get
\begin{equation} \langle\rho\rangle_{S^2}=4\pi\nu
Tr{\bf F}\wedge{\bf F}=Tr{\bf J} \label{rhona}
\end{equation}
This last expression shows that the particles are attached to the
``instanton density", which is topologically
conserved\footnote{This can also be written locally as $Trd({\bf
A}\wedge d {\bf A} - 2i/3{\bf A} \wedge {\bf A} \wedge {\bf A})$
but not always globally. Topologically non-trivial gauge
configurations on $S^4$ are classified by the homotopy group
$\pi_4(SU(2))=Z_2$. The integral of (\ref{rhona}) over the
four-sphere is generally non-vanishing. This is because the
integral of the Chern-Simons $3$-form over an equator three-sphere
may change under gauge transformations. This change is
proportional to some integer $n$. The instantons are classified by
the homotopy group $\pi_3(SU(2))=Z$.}.

The vortex free fluid is characterized by the unit instanton
number density
\begin{equation}
\int_{S^{4}} Tr{\bf F} \wedge {\bf
F}=\frac{16\pi^2(2I+1)I(I+1)}{3}
\end{equation}
Vortices in the fluid can change the instanton number by an
integer. These vortices are localized on the $4$-sphere. From the
six-dimensional point of view they are membranes wrapped on the
internal isospin sphere. We can also think of them as point-like
particles with many internal degrees of freedom.

Therefore, this section fully establishes the exact equivalence
between the CS theory in $6+1$ dimensions, defined over the $CP_3$
spatial manifold, and the CS theory in $4+1$ dimensions, defined
over the $S^4$ spatial manifold. In a previous section, we also
showed that the single particle equations of motion derived from
both formulations agree exactly with each other.

We conclude this section with some speculations. In the
thermodynamic limit (large $I$ limit), the $SU(2)$ representation
matrices are large. This limit is similar to the large $N$ limit
of QCD \cite{hooft} in which the leading contributions arise from
planar Feynman diagrams. The filling factor $\nu$ behaves as a 't
Hooft coupling constant. In this limit, the theory is strongly
coupled and should exhibit confinement, with strings or magnetic
flux tubes joining instanton-like vortices. Perhaps our theory is
dual to a topological string theory. Such a connection between
supersymmetric large $N$ CS theory in $2+1$ and topological string
theory was studied recently in \cite{vafa1}\cite{vafa2}. The
confining phase maybe thought of as a superconducting phase of the
liquid in the sense similar to the CSLG theory \cite{zhang1989}.
If this is the case, the QCD-like strings can be thought of as the
magnetic partners of the topological membranes we found in the
previous section. In $6+1$ dimensions the $3$-form coupling
electrically to membranes is dual to a $2$-form that naturally
couples to strings. We would like to speculate that the
worldvolume dynamics of these objects may lead to a `spin-gap' in
the spectrum of boundary excitations \cite{hu2001}. The finite
tension for the strings could arise from higher order corrections
at the magnetic length scale $I/R^2$.

\section{Fractional statistics of extended objects}
In this section, we describe the statistics of the membranes
constructed in a previous section. First, however, let us recall
the basic setup in the usual $2+1$ dimensional case.

In $2+1$ dimensions, particles interacting with velocity dependent
forces can become anyons with fractional statistics.  In $2+1$
dimensions we can obtain a topological current out of an $O(3)$
sigma-model field $n_a$ by virtue of:
\begin{eqnarray} J^\mu = \epsilon^{\mu \nu \lambda} \epsilon_{a b
c} n^a
\partial_\nu n^b \partial_\lambda n^c \end{eqnarray}
\noindent We exhibit the first Hopf ($S^3 \rightarrow S^2$) map by
$n^a/r = \bar {u} \sigma^a u$ where $u$ is given by (\ref{z}). The
conservation of $J^\mu$ allows us to obtain a gauge potential by
the equation:
\begin{eqnarray} \label{1formcurrentmanufactured} J^\mu =
\epsilon^{\mu \nu \lambda} \partial_\nu A_\lambda
\end{eqnarray}
\noindent The field $A$ is defined up to the usual abelian gauge
freedom. Then the homotopy invariant associated to the $S^3
\rightarrow S^2$ map is called the first Hopf invariant and is
given by \cite{Wilczek+Zee}
\begin{eqnarray} H_1 = - \int d^3 x A_\mu J^\mu \label{H1} \end{eqnarray} \noindent The
proof that $H_1$ is a homotopic invariant of the first Hopf map is
given in \cite{Wilczek+Zee}. Therefore, the action for the
solitonic object accepts added, in general, a topological term
$S=\theta H_1$ where we take $\theta = 2\pi \nu$, with $\nu$ being
the filling factor. The factor of $\theta$ reflects the fractional
charge of the soliton.  Rotating a soliton adiabatically (through
$2\pi$) over the time $T$, the wave-function acquires a phase
factor $\exp (iS_{total})$ where $S_{total}$ is the action
corresponding to the adiabatic rotation \cite{Wilczek+Zee}.
Typically, all other terms but the Hopf invariant will be of order
$1/T \rightarrow 0$ as $T \rightarrow \infty$.  The phase is given
by
\begin{eqnarray} \exp (i\Delta \phi)=\exp (i S) =\exp (2i \pi
\nu H_1) = \exp (2i \pi \nu)
\end{eqnarray} \noindent
with the Lagrangian and current given in (\ref{H1}) and
(\ref{1formcurrentmanufactured}).

There is a deep theorem which equates the Hopf invariant to the
linking number between two curves in $R^3$. The fractional nature
of the soliton can be visualized by using the linking number
theorem \cite{Hilton}. In $2+1$ dimensions, the world-lines
described by the particles can link and the particles can have
fractional spin and statistics. The relevant mathematics which
allows this is the homotopy $\pi_3(S^2)=Z$.

Parallel to the above discussion, in the $6+1$ dimensional case,
$J_{\Gamma\Gamma_1\Gamma_2}$, the conserved current defined in the
previous section, manufactures a 3-form field $C_{\Gamma \Gamma_1
\Gamma_2}$ through the curl equation
\begin{eqnarray} J^{\Gamma \Gamma_1 \Gamma_2} = \epsilon^{\Gamma
\Gamma_1 \Gamma_2 \Gamma_3 \Gamma_4 \Gamma_5 \Gamma_6}
\partial_{\Gamma_3} C_{\Gamma_4 \Gamma_5 \Gamma_6} \label{3formand5form} \end{eqnarray} \noindent
We exhibit the second Hopf map $S^7 \rightarrow S^4$ by
$X^a/R=\bar{\Psi} \Gamma^a \Psi$ where $\Gamma^a$ satisfy the
Clifford algebra. Then the homotopy invariant associated to the
second Hopf map is called the second Hopf invariant and is given
by: \begin{eqnarray} H_2=-\int d^7 x C_{\Gamma \Gamma_1 \Gamma_2}
J^{\Gamma \Gamma_1 \Gamma_2}
\end{eqnarray} \noindent This is associated with the homotopy group
$\pi_7(S^4)=Z \oplus Z_{12}$, where $Z_{12}$ is the torsion part
of the group. The action for the solitonic membranes accepts, in
general, a term $S= \theta H_2$ where $\theta = 2\pi \nu$, with
$\nu$ being the filling factor. A generalization of the theorem
which equates the Hopf invariant to the linking number between two
curves in $R^3$ guarantees the equivalence of $H_2$ with the
linking integral between two thin-membrane world-surfaces. The
phase acquired by taking one such (solitonic) membrane around the
other is \begin{eqnarray} \exp (i\Delta \phi)=\exp (i S) =\exp (i
2\pi \nu H_2) = \exp (i 2\pi \nu)
\end{eqnarray} \noindent with the Lagrangian and current as above. Our
membranes therefore acquire fractional statistics in $6+1$
dimensions by way of the Hopf invariant of the second Hopf map.
 We also mention that $H_1$ and $H_2$ with the $1$-form and
$3$-form gauge fields manufactured out of the conserved currents
via (\ref{1formcurrentmanufactured}) and (\ref{3formand5form}) are
precisely the terms in the action which one gets by having a
particle/membrane with fractional charge interacting with a
Chern-Simons gauge field.

Let us now give a simple example of how this interacting
materializes in our theory. Consider a thin static membrane
oriented along the $5-6$ plane, located at the origin in the other
$4$ directions. (See figure \ref{fig1}).  In this case,
$J^{056}=\delta^4(y)$ and
\begin{eqnarray} F_{1234} = -\delta^4(y) \label{createdfield}
\end{eqnarray} \noindent
Now consider a second thin membrane, perpendicular to the one we
already have, oriented along the $3-4$ plane at $y^1=0$ and
$y^2=c$, where $c$ is an arbitrary distance from the origin along
the $2$-axis. Move this membrane around the first one along a
circle in the $1-2$ plane (See Figure{\ref{fig1}). As it orbits
once, it describes a cylinder, $R^2_{3-4} \times S^1$. This is the
boundary of a four dimensional surface, $R^2_{3-4} \times D_2$
where $D_2$ is a disk in the $1-2$ plane. This membrane interacts
with the field in Eq.(\ref{createdfield}) produced by the first.
This interaction causes it to pick up a phase. Writing down the
Hopf invariant
\begin{displaymath} \exp(i
\delta \phi) = \exp(-2i \pi \nu \int d^7 x J \cdot C) = \exp(-2i
\pi \nu\int_{R^2_{3-4} \times S^1} C)=
\end{displaymath}
\begin{eqnarray}= \exp(-2i \pi \nu \int_{R^2_{3-4} \times D_2}  F )= \exp(2\pi i \nu)
\end{eqnarray} \noindent \noindent where $C$ is the $3$-form gauge field
defined in Eq.(\ref{3formand5form}). This phase corresponds to an
anyonic exchange phase of \begin{eqnarray} \delta \phi =\pi \nu
\end{eqnarray} \noindent

\section{Conclusions}
From this work we see that the precise connection between the microscopic
wave function and the CSLG topological field theory description of the
2D quantum Hall effect can be directly generalized to the new quantum
liquid constructed by ZH. The abelian $U(1)$ CS theory in 6+1 dimensions
and the $SU(2)$ non-abelian CS theory in $4+1$ dimensions can be constructed
directly by a proper generalization of the flux attachment transformation.
This effective field theory model can be used to investigate many long
wave length properties. In particular, it would be interesting to
study the new quantum Hall liquid on different topological backgrounds.
It would also be interesting to apply this
topological field theory to study the boundary excitations. The new quantum
liquid also supports topologically stable extended objects, including the
membrane and the 4-brane, whose dynamics is completely determined by the
topological action. The 2 branes can intersect each other in a non-trivial
way to give rise to the fractional statistics in higher dimensions.
The precise microscopic connection should help us to develop a fully
regularized quantum theory of these extended objects. The world volume
dynamics of these extended objects, especially the membrane, could
lead to gaps to higher helicity states. In this case, low energy
isospin degrees would be finite, and a truly four dimensional theory
could be obtained.

We would like to acknowledge useful discussions with M. Fabinger,
D. Oprea, A. Polyakov, D. Santiago, L. Susskind and F. Wilczek. NT
would like to thank N. Arkani-Hamed, S. Minwalla, L. Motl, S. Nam,
A. Neitzke, and B. Pioline. CHC is grateful to W.-Y. Chuang for
stimulating discussions.  This work is supported by the NSF under
grant numbers DMR-9814289. J.P. Hu is also supported by the
Stanford Graduate fellowship.  The work of NT is supported by the
DOE, grant number DE-FG02-91ER40654.

\section{Appendix}

\subsection{Notations and conventions}
$A,B=1,2,3,4,5,6$ denote the local $CP_3$ coordinates.

$\Gamma,\Sigma=0,1,2,3,4,5,6$ denote space-time coordinates on $CP_3$.

$a=1,2,3,4,5$ denote the $S^4$ coordinates.

$i=1,2,3$ denote the $S^2$ coordinates.

${\bf A}_a$, ${\bf F}_{ab}$ and ${\bf I}_i$ denote $SU(2)$ matrix
valued quantities.

${\cal A}_B$ and ${\cal F}_{AB}$  denote the total gauge potential
and field strength.

$A_B$ and $F_{AB}$  denote the background gauge potential and
field strength.

$a_B$ and $f_{AB}$  denote the fluctuating part of the gauge
potential and field strength.

\subsection{Local metric on $CP_3 \sim S^{4}\times S^{2}$}
We can parameterize the 2-sphere and the 4-sphere as follows.
For $S^{2}$, we define
\begin{eqnarray} \nonumber
& & n_{1}\!=\! r\sin{\alpha}\cos{\beta}   \\ \nonumber & &
n_{2}\!=\! r\sin{\alpha} \sin{\beta}  \\ & & n_{3}\!=\!
r\cos{\alpha} \label{s2}
\end{eqnarray}
For $S^{4}$, we define
\begin{eqnarray} \nonumber
& &X_{1}\!=\! R\sin{\theta_{1}}\sin{\frac{\theta_{2}}{2}}\sin{(\phi_{1}-\phi_{2})} \\
\nonumber & &X_{2}\!=\! -R\sin{\theta_{1}}\sin{\frac{\theta_{2}}{2}}\cos{(\phi_{1}-\phi_{2})}  \\
\nonumber & &X_{3}\!=\! -R\sin{\theta_{1}}\cos{\frac{\theta_{2}}{2}}\sin{(\phi_{1}+\phi_{2})}  \\
\nonumber & &X_{4}\!=\! R\sin{\theta_{1}}\cos{\frac{\theta_{2}}{2}}\cos{(\phi_{1}+\phi_{2})} \\
& &X_{5}\!=\! R\cos{\theta_{1}} \label{s4}
\end{eqnarray}
$r$ and $R$ are kept constant.  These are the spherical
coordinates for the 2-sphere and 4-sphere respectively.  The
metrics on these spheres are given by
\begin{eqnarray}
ds^{2}=r^{2}d\omega^{2}, \ \ \ \
d\omega^{2}=d\alpha^{2}+\sin^{2}{\alpha} d\beta^{2}
\label{s2metric}
\end{eqnarray}
\begin{eqnarray}
dS^{2}=R^{2}d\Omega^{2}, \ \ \ \
d\Omega^{2}=d\theta^{2}_{1}+\frac{\sin^{2}{\theta_{1}}}{4}d\theta^{2}_{2}+\sin^{2}{\theta_{1}}d\phi^{2}_{1}+\sin^{2}{\theta_{1}}d\phi^{2}_{2}
\label{s4metric} \end{eqnarray}

\subsection{K\"ahler form on $CP_n$}
We will follow closely the treatment of Greene \cite{greene}.
$CP_{n}$ is defined by introducing $n+1$ complex coordinates $z_1,
\ldots ,z_{n+1}$, not all of them simultaneously zero, with an
equivalence relation identifying  $z_1, \ldots ,z_{n+1}$ with
$\lambda z_1, \ldots ,\lambda z_{n+1}$ for any complex number
$\lambda$ other than zero. With a suitable choice of $|\lambda|$
we can always choose the representatives of such an equivalence
class to satisfy
\begin{equation}
\sum_{i}\bar{z}_i z_i =1 \label{modulus}
\end{equation}
This condition fixes only part of the equivalence relation
defining $CP_n$. We still have to impose identifications
associated with the phase of $\lambda$
\begin{equation}
(z_1, \ldots, z_{n+1}) \sim e^{i\theta}(z_1, \ldots, z_{n+1})
\end{equation}
We introduce local coordinates as follows. Define the $j$-th patch
$X_j$ by the condition that $z_j \ne 0$ and set $z_j = s_j
e^{i\alpha_j}$ with $s_j$ real. Then define
\begin{equation}
(s_j,\ u_{(j)}^1, \ldots , u_{(j)}^n)=(s_j,\ e^{-i\alpha_j}z_1,
\ldots e^{-i\alpha_j}z_{j-1},\ e^{-i\alpha_j}z_{j+1}, \ldots ,
e^{-i\alpha_j}z_{n+1}) \label{fixequi}
\end{equation}
to be local coordinates on the patch. Using (\ref{modulus}) one
can solve for $s_j$ in terms of the $u_{(j)}$'s and work with $n$
independent complex coordinates. We define the K\"ahler scalar
potential to be $K_j= 2\sum_1^n |u^{i}_{(j)}|^2$. Then the
$2$-form
\begin{equation}
J=\partial \bar{\partial}K_j=dA_j, \,\
A_j=2s_jds_j+2\sum_1^n\bar{u}^{i}_{(j)}du^{i}_{(j)}
\end{equation}
is a {\it globally} defined closed $2$-form class on $CP_n$ and
equivalently it defines a K\"ahler metric. To see that $J$ is
globally defined, all we must check is that in two overlapping
patches $X_{j_1},\ X_{j_2}$ the vector potentials $A_{j_1}$ and
$A_{j_2}$ are related by a gauge transformation. But by
construction $A_j=2\sum_1^{n+1}\bar{z}^{i}dz^i-2d\alpha_j$ and so
$A_{j_1}=A_{j_2}+2d(\alpha_{j_{2}}-\alpha_{j_1})$.

We now consider the two examples of interest to us, namely,  the
sphere $CP_1$ and $CP_3$ and establish a {\it local} equivalence
of the background field strength $F$ with $J$. For simplicity we
demonstrate the case of $CP_1$ explicitly. For $CP_1$, let us work
in a patch such that $z_1 \ne 0$ and use equation (\ref{z}).  We
take $u_1$ to be real as in (\ref{fixequi}) and parameterize the
$u$'s using cartesian coordinates $n_i$ as in (\ref{u}). The south
pole $n_3=-r$ is excluded from the patch. Then, {\it locally}
\begin{equation}
J=2d\bar{u} \wedge du=-{i \over 2r^3}\epsilon_{ijk}n_k dn_i \wedge
dn_j=-iF
\end{equation}
where $F$ is the field strength of a $U(1)$ magnetic monopole at
the center. Similarly for $CP_3$, we can work in the patch $z_1
\ne 0$ and set $(\Psi_1,\ldots , \Psi_4)$ to be given by
(\ref{Psi}) in terms of the $X_a, \ n_i$. The $2$-sphere $X_5=-R$
together with the $4$-sphere $n_3=-r$ are excluded from the patch.
Then {\it locally}
\begin{equation}
J= 2d\bar{\Psi} \wedge d\Psi=-iF
\end{equation}
with $F$ given explicitly by (\ref{F}). This last equation is
equivalent to the Berry's phase computation appearing in
(\ref{single}).

\subsection{Useful identities of A and F}
In this section, we summarize some careful results for deriving
the single particle equation of motion\footnote{We also set
$I=\frac{1}{2}$ here.} (\ref{eomu1}).
\begin{eqnarray}
A_{a}A_{a}&=&A_{\mu}A_{\mu}=-\frac{R-X_{5}}{4R^{2}(R+X_{5})} \\
F_{5a}F_{a5}&=&\frac{R^{2}-X^{2}_{5}}{4R^{6}} \\
F_{5\mu}F_{\mu\nu}&=&\frac{X_{\nu}X_{5}}{4R^{6}} \\
F_{\mu
i}F_{ik}&=&\frac{-1}{4r^{4}R(R+X_{5})}\varepsilon_{ikl}\eta^{i}_{\mu\nu}n_{l}X_{\nu}
\\
F_{\mu\tau}F_{\tau\nu}&=&\frac{\delta_{\mu\nu}}{R^{2}(R+X_{5})^{2}}+\frac{(2R+X_{5})(2R+3X_{5})}{R^{2}(R+X_{5})^{2}}A_{\mu}A_{\nu}-\frac{(2R+X_{5})^{2}}{4R^{6}(R+X_{5})^{2}}X_{\mu}X_{\nu}
\\
F_{ij}F_{jk}&=&\frac{1}{4r^{4}}(\delta_{ik}-\frac{n_{i}n_{k}}{r^{2}})\\
\sum_{a\neq b}&F_{ab}&F_{ab}+2\sum_{a,i}F_{ai}F_{ai}+\sum_{i\neq j
}F_{ij}F_{ij}\\ \nonumber
&=&-\frac{3R^{4}+R^{2}X^{2}_{5}+4RX^{3}_{5}-2X_{5}R^{3}+2X^{4}_{5}}{2R^{6}(R+X_{5})^{2}}-\frac{R-X_{5}}{r^{2}R^{2}(R+X_{5})}-\frac{1}{2r^{4}}
\end{eqnarray}

\bibliographystyle{prsty}
%\bibliography{cs}

\begin{figure*}[ht]
\includegraphics{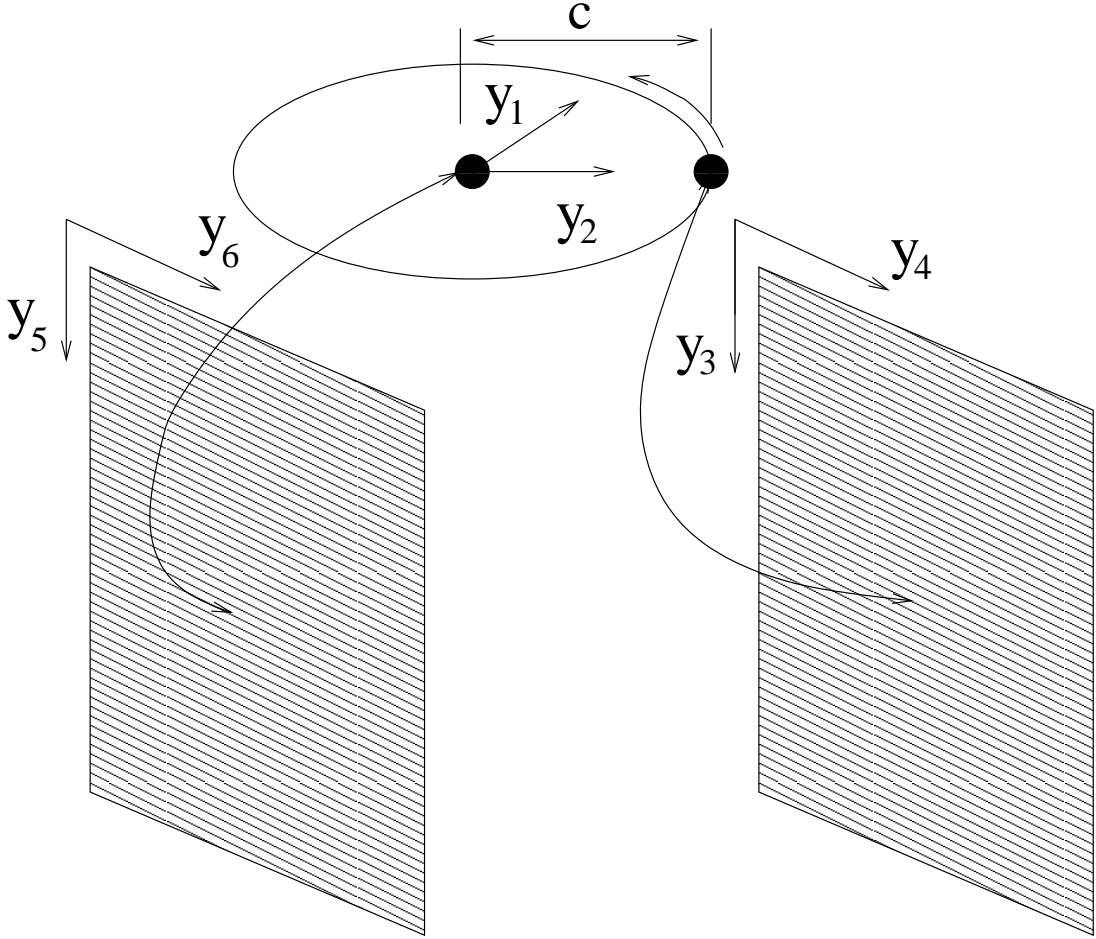}
\caption{The thin membrane on the $3$ - $4$ directions, originally
at a distance $c$ along the $2$-axis from the thin static membrane
on the $5$ - $6$ directions is rotated around the latter along a
circle in the $1$-$2$ plane. After rotation, it picks up a
fractional phase.} \label{fig1}
\end{figure*}

\end{document}